\documentclass[twocolumn]{aastex702}
\usepackage{placeins}

\usepackage{graphicx}
\usepackage{booktabs}
\shorttitle{Dark Matter Cusp in a Pop III.1 Minihalo}
\shortauthors{Bhowmick et al.}

\submitjournal{ApJL}

\begin{document}

\title{
Emergence of a Steep Dark Matter Cusp at the Onset of Population III.1 Star Formation: Implications for Supermassive Black Hole Seeding
}

\author[orcid=0009-0004-0113-4550,
gname=Sagnik,sname=Bhowmick]{Sagnik Bhowmick}
\affiliation{Department of Physical Sciences,
Indian Institute of Science Education and Research Kolkata,
Mohanpur, Nadia 741246, India}
\email{sb23ms007@iiserkol.ac.in}

\author[orcid=0000-0002-3389-9142,
gname=Jonathan,sname=Tan]{Jonathan C. Tan}
\affiliation{Department of Physics and Astronomy,
Chalmers University of Technology,
SE-412 96 Gothenburg, Sweden}
\affiliation{Department of Astronomy,
University of Virginia,
Charlottesville, VA 22904, USA}
\email{jonathan.tan@chalmers.se}

\author[orcid=0000-0002-4317-767X,gname=Shingo,sname=Hirano]{Shingo Hirano}
\affiliation{Department of Applied Physics, Faculty of Engineering,
Kanagawa University, Kanagawa 221-0802, Japan}
\affiliation{Department of Astronomy, School of Science,
University of Tokyo, Tokyo 113-0033, Japan}
\email{shingo-hirano@kanagawa-u.ac.jp}

\author[orcid=0000-0001-7925-238X]{Naoki Yoshida}
\affiliation{Department of Physics, School of Science,
The University of Tokyo, 7-3-1 Hongo, Bunkyo,
Tokyo 113-0033, Japan}
\affiliation{Kavli Institute for the Physics and Mathematics
of the Universe (WPI), UT Institutes for Advanced Study,
The University of Tokyo, 5-1-5 Kashiwanoha,
Kashiwa, Chiba 277-8583, Japan}
\email{naoki.yoshida@phys.s.u-tokyo.ac.jp}

\begin{abstract}
An enhanced central dark matter cusp is needed for scenarios in which weakly interacting massive particle (WIMP) self-annihilation heating allows extended growth of Population (Pop) III.1 protostars to produce supermassive black hole (SMBH) seeds. Such cusps are expected to vary among Pop III minihalos. In the Pop III.1 scenario of Banik, Tan \& Monaco (2019) it is only expected to occur in very rare first-forming minihalos that are locally isolated from astrophysical sources of ionizing feedback, where dynamically slow baryonic collapse leads to adiabatic contraction of the central dark matter distribution. We simulate such a Pop III.1 minihalo using multiresolution cosmological zoom-in simulations, including nonequilibrium primordial thermochemistry. We find that the central dark matter profile steepens progressively as the gas collapses. In our highest-resolution simulation, at a peak
gas density of $10^{12}\,\mathrm{GeV\,cm^{-3}}$, the inner profile reaches a power-law index of $\alpha=1.933\pm0.153$, where
$\rho_\chi\propto r^{-\alpha}$, measured outside the adopted gravitational-softening limit.
Extrapolating this profile to a characteristic protostellar radius of $100\,R_\odot$ yields a dark matter density of $1.1\times10^{14}\,\mathrm{GeV\,cm^{-3}}$, which we expect to be a lower limit on the eventual density when the star first forms. For standard rates of WIMP self-annihilation, such densities are expected to have a significant impact on the thermodynamics of the star-forming gas. These densities may also enable high rates of WIMP capture by the Pop III.1 protostar that are essential for its growth to $\gtrsim10^4\:M_\odot$.
\end{abstract}


\section{Introduction}
\label{sec:intro}

The discovery of high-redshift quasars hosting supermassive black holes (SMBHs)
\citep[e.g.,][]{Mortlock2011,Wang2021} challenges models of early black hole
seeding and growth. Reaching $\gtrsim10^9\,M_\odot$ within the first billion years
from relatively low-mass ($\sim100\:M_\odot$) black holes that are the expected remnants from traditional models of Pop III star formation \citep[e.g.,][]{2002Sci...295...93A,2002ApJ...564...23B,2004ApJ...603..383T,2008ApJ...681..771M} requires efficient, even super-Eddington, accretion to be sustained for several hundred Myr. Such accretion rates appear to be difficult to realize in the presence of star formation and active galactic nucleus (AGN) feedback \citep[e.g.,][]{2007MNRAS.374.1557J,2009ApJ...698..766M,2025MNRAS.544.4317S}, which thus motivates scenarios for ``heavy'' initial black hole seed masses that are already in or near the supermassive regime, i.e., $\gtrsim10^4\:M_\odot$. The apparent dearth of intermediate-mass black holes (IMBHs) in the local universe \citep[e.g.,][]{2020ARA&A..58..257G,2025MNRAS.541..429M} suggests that such heavy seeds may be a characteristic feature of the
entire SMBH population, which is estimated to have cosmic abundance of at least $\sim10^{-2}\:{\rm cMpc}^{-3}$, both locally and out to redshifts $z\sim7$ \citep[][]{2024ApJ...971L..16H,2025ApJ...991..141C}.

Most physically-motivated heavy seed models have difficulties in reconciling these features. For example, ``direct collapse black holes'' (DCBHs) that arise from irradiated, metal-free atomically-cooled halos, i.e., with masses $\sim10^7-10^8\:M_\odot$, have typical abundances $\lesssim10^{-4}\:{\rm cMpc}^{-3}$ \citep[e.g.,][]{2016ApJ...832..134C,2024MNRAS.531.4584S,2025OJAp....8E..88O}. 
Models of these halos in which gas cooling and fragmentation are suppressed by turbulent dynamical heating, rather than external radiation, predict even lower abundances $\lesssim10^{-6}\:{\rm cMpc}^{-3}$ \citep[][]{2019Natur.566...85W,2022Natur.607...48L}. Models invoking runaway stellar collisions in dense clusters \citep[e.g.,][]{2006MNRAS.368..141F,Sakurai17,2026MNRAS.549ag986R} require extreme stellar densities that are rarely observed, at least in the local universe \citep[e.g.,][]{2014prpl.conf..149T}. Furthermore, such collisional growth scenarios would be expected to produce very large numbers of IMBHs, which, as mentioned, have not yet been observed. 

An alternative way to produce abundant heavy SMBHs in the early universe is presented in the Pop III.1 seeding model \citep{Banik2019,2023MNRAS.525..969S,2025MNRAS.536..851C,2026arXiv260528777P} \citep[see][for a review]{2024arXiv241201828T}. In this scenario, the physical process of weakly interacting massive particle (WIMP) self-annihilation heating \citep{Spolyar2008,Natarajan2009} is invoked to impact the dynamics of gas collapse to the forming star, especially reducing fragmentation \citep{2012ApJ...761..154S,2014MNRAS.441..822S}, and the protostellar evolution, especially delaying contraction to the main sequence that is critical for growth to large masses \citep{2008ApJ...685L.101F,2011ApJ...736...58H,
2015ApJ...799..210R,Nandal2025,Topalakis2025}.
Some attractive features of the Pop III.1 seeding model are that it predicted a large abundance of SMBHs at high redshifts in advance of observational discoveries with the James Webb Space Telescope (JWST) \citep[e.g.,][]{2023ApJ...959...39H} and the Hubble Space Telescope (HST) \citep[][]{2024ApJ...971L..16H,2025ApJ...991..141C}. It also presents a physically-motivated mechanism to create the entire population of SMBHs with an abundance of $\sim0.1\:{\rm cMpc}^{-3}$ \citep{2024arXiv241201828T,2025MNRAS.542.1532S,2026arXiv260528777P} based on R-type HII region expansion into the intergalactic medium at $z\sim20$ driven by Pop III.1 supermassive stars, with further SMBH seeding in these regions suppressed due to the expected fragmented collapse of irradiated Pop III.2 minihalos \citep{2005ApJ...628L...5O,2006MNRAS.373..128G,Yoshida07}. Intriguingly, the predicted Pop III.1 ``flash'' ionization of the universe raises the Thomson scattering optical depth of CMB photons by up to $\sim50\%$ \citep{2025ApJ...989L..47T,2026ApJ..1005L..32T}, with such a boost independently invoked to solve several tensions facing $\Lambda$CDM in the form of dynamical dark energy and preference for negative neutrino masses \citep{Jhaveri2025,Sailer2026}.


For self-annihilating dark matter particles of mass $m_\chi$ and thermally averaged annihilation cross section $\langle\sigma v\rangle$, the local energy deposition rate is
\begin{equation}
\Gamma_\chi = f_{\rm heat}
\frac{\langle\sigma v\rangle}{m_\chi}\rho_\chi^2c^2,
\label{eq:heating}
\end{equation}
where $\rho_\chi$ is the dark matter mass density and $f_{\rm heat}$ is the fraction of annihilation energy deposited locally in the gas. The explicit $c^2$ is omitted when $m_\chi$ and $\rho_\chi$ are
expressed in energy units. Because $\Gamma_\chi\propto\rho_\chi^2$,
modest changes in central dark matter density can strongly affect the local heating rate in the protostellar environment. The rate of capture of WIMPs by a protostar is also sensitive, linearly, to the local dark matter density. Whether the required densities for there to be a significant impact on protostellar fragmentation and evolution develop naturally during primordial collapse is therefore a central
question for the Pop III.1 SMBH formation channel.

Pop III.1 stars form from metal-free gas in minihalos that have not been significantly affected by earlier stellar feedback
\citep{Banik2019}.
Cooling by H$_2$ and, under suitable conditions, HD allows the gas
to condense.
As primordial gas cools and concentrates near a halo's center, its gravity modifies dark matter orbits through adiabatic contraction. In the classical circular-orbit approximation, increasing enclosed baryonic mass shrinks the orbits \citep{Blumenthal1986}. Real minihalos contain noncircular orbits and evolving, often nonspherical potentials, and their response depends on halo assembly, orbital distributions, and the spatial and temporal evolution of the gas. Thus, gas collapse does not necessarily produce a steep dark
matter cusp.

Cosmological calculations have previously found central dark matter enhancements during primordial star formation \citep{Natarajan2009}. Using high-resolution simulations, \citet{HiranoYoshida2025} found that their fiducial minihalo, hereafter HaloHY, developed a comparatively shallow inner
profile, $\rho_\chi\propto r^{-0.6}$, while other minihalos showed a range of central slopes. This variation motivates examining how steeper cusps emerge and evolve during collapse.

Here we follow the collapse of a minihalo that is selected to be representative of Pop III.1 sources, i.e., it is the first to form in its local region.
We examine the evolution of its central density and slope, the radial structure revealed at increasing resolution, and the implications of continuing the fitted cusp to protostellar scales.

\section{Numerical Methods}
\label{sec:methods}

We simulate the formation and gravitational collapse of primordial gas clouds with \textsc{GADGET3-DAWN}, following \citet{HiranoYoshida2025}. The code couples collisionless $N$-body dynamics and smoothed-particle hydrodynamics (SPH) in a self-consistent gravitational potential, with nonequilibrium
primordial chemistry and atomic and molecular cooling and heating, including H$_2$ and HD \citep{Yoshida08}. We assume metal-free gas, no imposed external Lyman--Werner background, and zero initial coherent baryon--dark matter streaming velocity ($v_{\rm bc}=0$). Dark matter annihilation heating is not directly coupled to the hydrodynamic evolution, but we assess its potential impact in a post-processing step using the simulated profiles.

We identify the target minihalo, hereafter HaloM, in a dark-matter-only simulation of a periodic box with side length $L_{\rm box}=1\,h^{-1}\,\mathrm{cMpc}$. The initial conditions are
generated with \textsc{MUSIC} \citep{HahnAbel2011} at $z_{\rm ini}=499$ for a flat $\Lambda$CDM cosmology with $\Omega_{\rm m}=0.31$,
$\Omega_{\rm b}=0.048$, $\Omega_\Lambda=0.69$,
$H_0=68\,\mathrm{km\,s^{-1}\,Mpc^{-1}}$, $n_{\rm s}=0.96$, and $\sigma_8=0.80$. After evolving the parent simulation to $z=20$, we use the \textsc{ROCKSTAR} halo finder \citep{Behroozi2013}
to select the most massive minihalo in the box.

At this \emph{selection epoch}, HaloM has
$M_{\rm vir}=9.55\times10^5\,h^{-1}M_\odot\simeq
1.40\times10^6\,M_\odot$, a virial radius
$R_{\rm vir}=2.46\,h^{-1}\,\mathrm{ckpc}$, scale radius
$r_{\rm s}=1.56\,h^{-1}\,\mathrm{ckpc}$, maximum circular velocity
$V_{\max}=5.96\,\mathrm{km\,s^{-1}}$, and catalogue and Bullock
spin parameters of $0.0183$ and $0.0267$, respectively. These catalogue
properties are measured at $z=20$; the gas-collapse snapshot
analysed below is at $z\simeq29.53$. We trace the selected halo particles back
to their Lagrangian region at $z=499$, construct refined zoom-in
initial conditions containing gas and dark matter, and evolve this
hydrodynamic realization, L0, from $z=499$ through primordial collapse.

We refine using the 13-child splitting prescription of \citet{HiranoYoshida2025}, with the refinement region determined recursively from the preceding run's collapse history. When L0 first reaches a peak gas density of approximately $10^6\,\mathrm{GeV\,cm^{-3}}$, we select gas particles above $10^2\,\mathrm{GeV\,cm^{-3}}$ and trace them back to the L0 snapshot at $z=49$. A sphere of radius $R_{\rm ref}$ encloses their progenitors; we add a 10\% radial buffer, i.e.,
$R_{\rm split}=1.1R_{\rm ref}$.
Within this sphere, each gas and dark matter parent particle is replaced by 13 equal-mass children in a compact, approximately symmetric configuration, conserving parent mass and linear momentum, while inheriting its bulk velocity. Evolving the refined state from $z=49$ yields L1. The same selection, backward tracing, and splitting procedure produces L2 and L3 recursively. All splitting occurs at $z=49$, before strong central collapse, i.e., not within an already contracted cusp.

For each refinement level $L=0,1,2,3$, particle masses and softening lengths are given by:
\begin{equation}
 m_{i,L}=m_{i,0}\,13^{-L},\qquad
 i\in\{\mathrm{gas},\mathrm{halo}\},
\label{eq:particle_mass}
\end{equation}
\begin{equation}
 \epsilon_{i,L}=\epsilon_{i,0}\,13^{-L/3}.
\label{eq:softening_scaling}
\end{equation}
Thus, L3 has 2197 times finer mass resolution and 13 times smaller softening than L0. Gas and dark matter particle masses change from $0.285$ and $1.556\,M_\odot$ in L0 to $1.30\times10^{-4}$ and $7.08\times10^{-4}\,M_\odot$ in L3. Comoving dark matter softening decreases from $4.883\,h^{-1}\,\mathrm{cpc}$ to $0.376\,h^{-1}\,\mathrm{cpc}$, equivalent to a physical softening of $\simeq0.0181\,\mathrm{pc}$ at the final stage.

To reduce the cost of the high-density evolution, we construct a local particle cutout at a peak gas density near $10^6\,\mathrm{GeV\,cm^{-3}}$, retaining gas and dark matter particles within $3R_{\rm vir}$ of the collapsing gas center. These particles are recentered and evolved with nonperiodic boundary conditions. A comparison against continued evolution in the original cosmological volume shows close agreement of the inner gas and dark matter density profiles over the tested interval. The highest-resolution calculation
reaches a peak gas density of approximately $10^{12}\,\mathrm{GeV\,cm^{-3}}$.

For each output, we define the center as the mean position of the ten highest-density gas particles and construct spherically averaged gas and dark matter densities and enclosed masses in radial shells. The inner dark matter fits use logarithmic bins with at least 100 halo particles per bin. We restrict the fits to radii satisfying
\begin{equation}
 r\geq r_{\min,\chi}
 =\max\left(r_{100,\chi},3.2\epsilon_{\chi}\right),
\label{eq:dm_resolution}
\end{equation}
where $r_{100,\chi}$ encloses 100 halo particles and $\epsilon_\chi$ is their gravitational softening length. At the final L3 output the softening-based boundary is $0.058\,\mathrm{pc}$. We represent the inner dark matter distribution by
\begin{equation}
 \rho_\chi(r)=\rho_0\left(\frac{r}{r_0}\right)^{-\alpha},
\label{eq:dm_powerlaw}
\end{equation}
where $\alpha$ is the positive logarithmic slope and $\rho_0$ the normalization at $r_0$. For the reported inner-cusp slopes, we examine admissible windows containing 10--26 radial bins and retain the window with the largest fitted $\alpha$; quoted uncertainties are formal fit errors. For the final HaloM L3 profile, the selected inner fitting interval spans approximately $0.058$--$0.086\,\mathrm{pc}$ and contains $14{,}007$ dark matter particles distributed across 10 logarithmically spaced radial bins. Each fitted bin contains at least 100 dark matter particles, satisfying our adopted minimum particle-count criterion.

This interval is selected by the stated steepest-admissible-window procedure above the adopted resolution boundary, rather than by an assumed physical transition radius. For the HaloM--HaloHY
comparison,
we instead use the common $0.068$--$0.185\,\mathrm{pc}$ interval over which both profiles are retained, so the power law indices are fitted over the same physical radii. We compare the profiles across the resolution sequence and successive L3 outputs and use the fitted inner power law to estimate densities at protostellar radii.

For the illustrative annihilation-heating comparison, we evaluate Equation~(\ref{eq:heating}) in post-processing for the final L3 gas
state, adopting $m_\chi=500\,\mathrm{GeV}$ and
$\langle\sigma v\rangle=3\times10^{-26}\,
\mathrm{cm^3\,s^{-1}}$. We adopt $f_{\rm heat}=0.1$ (fiducial), $1/3$, and $2/3$ (optimistic) \citep{2014MNRAS.441..822S}. Denoting the estimated local heating by $H(r)\equiv\Gamma_\chi(r)$, we compare it with $C_{\rm net}(r)$, the net gas microphysical cooling rate per unit volume: cooling minus heating from the gas source terms, excluding compression, shock heating, and dark matter annihilation. For $C_{\rm net}>0$, we define the local heating-to-cooling ratio as
\begin{equation}
\mathcal{R}_\chi(r)=\frac{\Gamma_\chi(r)}{C_{\rm net}(r)},
\label{eq:heating_cooling_ratio}
\end{equation}
and evaluate it using the final gas state. The dark matter density is taken from the measured profile at resolved radii and from its fitted inward power-law continuation at smaller radii. The calculation does not transport annihilation products or include annihilation heating in the simulated gas evolution.

\section{Results}
\label{sec:results}

We examine gas collapse and the dark matter response in HaloM, followed by resolution dependence, a comparison with HaloHY, and evolution of the inner cusp.

\begin{figure*}[t]
\centering
\includegraphics[width=0.96\textwidth]{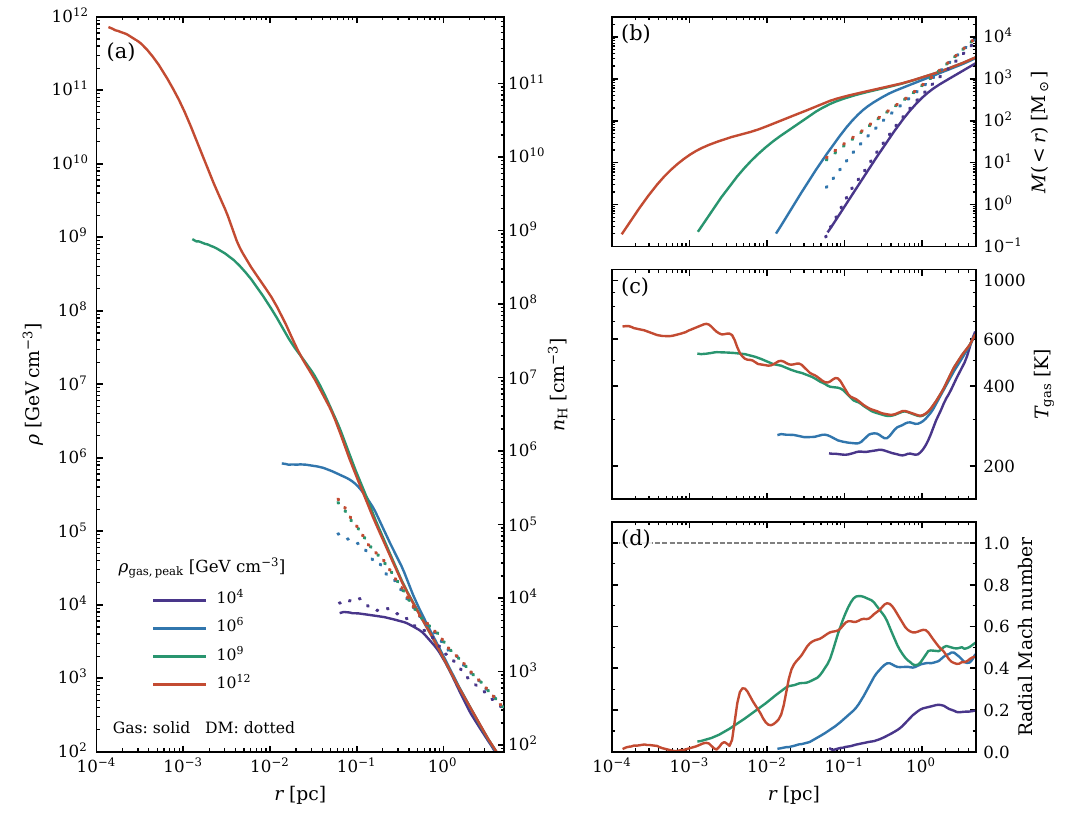}
\caption{Evolution of HaloM during primordial gas collapse in the L3 simulation, at peak gas densities of $10^4$, $10^6$, $10^9$, and $10^{12}\,\mathrm{GeV\,cm^{-3}}$. \textbf{(a)} Spherically averaged gas and dark matter density profiles, with the corresponding gas hydrogen-number-density scale
shown on the secondary axis. \textbf{(b)} Enclosed gas and dark matter masses at the same stages. \textbf{(c)} Gas temperature profiles.
\textbf{(d)} Radial inflow Mach number,
$\mathcal{M}_r=|\langle v_r\rangle|/\langle c_s\rangle$; the grey dashed horizontal line marks unity. Purple, blue, green, and red identify successively higher peak gas densities in every panel. Gas is shown by solid lines and dark matter
by dotted lines in panels (a) and (b).}
\label{fig:halom_radial_evolution}
\end{figure*}

Figure~\ref{fig:halom_radial_evolution} follows the L3 calculation at peak gas densities of $10^4$, $10^6$, $10^9$, and $10^{12}\,\mathrm{GeV\,cm^{-3}}$.
For primordial gas with hydrogen and helium mass fractions
of $X_{\rm H}=0.76$ and $Y_{\rm He}=0.24$, respectively,
the hydrogen nuclei number density is
$n_{\rm H}/\mathrm{cm^{-3}}\simeq
0.81\,\rho_{\rm gas}/(\mathrm{GeV\,cm^{-3}})$,
where $\rho_{\rm gas}$ includes the contribution from helium. Panel (a) shows that gas becomes strongly concentrated at small radii while its outer profile changes little. The dark matter responds more gradually, with increased density
at smaller resolved radii. Its profiles at peak gas densities of $10^9$ and $10^{12}\,\mathrm{GeV\,cm^{-3}}$ nearly overlap despite continued gas collapse. Either most detectable contraction at these radii occurred earlier, or further dark matter contraction is confined below the adopted resolution boundary. The present profiles do not distinguish these possibilities.

The enclosed-mass curves in panel (b) make this redistribution clearer. Gas accumulates at progressively smaller radii, increasing the mass
contained in the central region. The dark matter enclosed mass also responds to the evolving gravitational potential. In the spherical approximation, the acceleration from each component is $g_i(r)=GM_i(<r)/r^2$, so the buildup of central baryonic mass provides a natural driver for the inward response of dark matter.

The thermal evolution does not simply follow a monotonic temperature increase. At the \(10^9\,\mathrm{GeV\,cm^{-3}}\) stage, panel (c) shows an inner gas temperature of several hundred K, alongside an already concentrated gas core. This is consistent with substantial central gas collapse; the temperature by itself is not a criterion for collapse of the entire halo. At the final stage, the inner gas is again at several hundred K, reflecting the competition among molecular cooling, chemical processes, and compressional heating. Meanwhile, the shell-averaged radial inflow Mach number in panel (d), $\mathcal{M}_r=|\langle v_r\rangle|/\langle c_s\rangle$, remains mostly below unity over the plotted radii. Thus, much of the central mass accumulation occurs through sustained subsonic inflow, even as the inner gas becomes highly concentrated.

\begin{figure*}[t]
\centering
\includegraphics[width=0.96\textwidth]{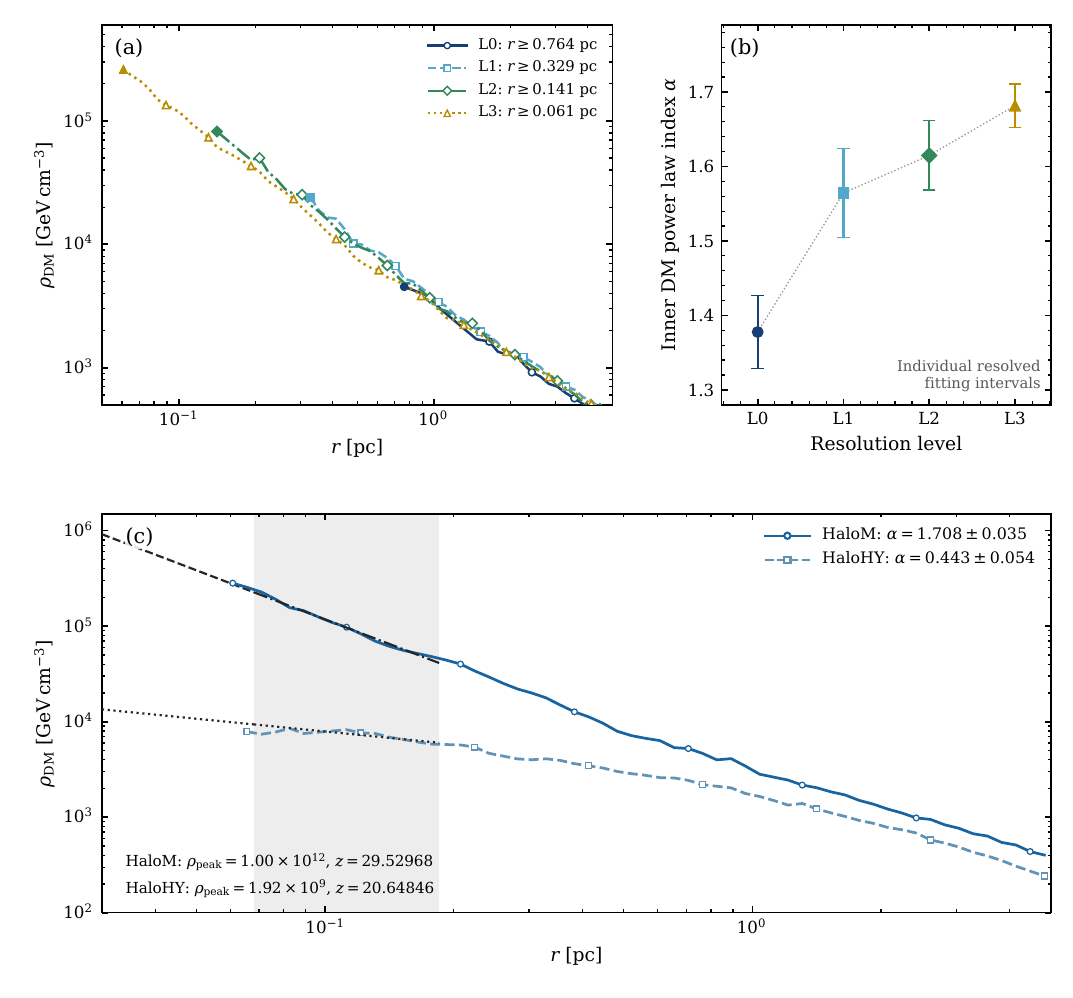}
\caption{Dependence of the dark matter cusp on numerical resolution and
comparison with \citet{HiranoYoshida2025}.
\textbf{(a)} Measured radial dark matter density profiles for HaloM L0--L3 at an approximately matched collapse stage. The measured peak gas densities are $(1.884,\,2.535,\,1.994,\,2.504)\times10^9\,
\mathrm{GeV\,cm^{-3}}$, and the first displayed shell centres are $0.764$, $0.329$, $0.141$, and $0.061\,\mathrm{pc}$, respectively.
\textbf{(b)} Inner power law indices fitted over the first 13 fully resolved shells at each resolution. The physical fitting intervals differ across levels.
\textbf{(c)} Dark matter density profiles of HaloM L3 and HaloHY, with power-law indices of $1.708\pm0.035$ and $0.443\pm0.054$, respectively, over the common shaded fitting interval of $0.068$--$0.185\,\mathrm{pc}$. The inward extensions of the fitted curves are extrapolations. 
}
\label{fig:halom_resolution_comparison}
\end{figure*}

The resolution sequence in Figure~\ref{fig:halom_resolution_comparison} provides another view of the inner structure. Panel (a) compares the L0--L3 dark matter profiles at the highest collapse stage common to all four calculations. The profiles agree reasonably well at larger shared radii, while refinement allows the inner region to be examined more closely. Their adopted minimum radii are approximately $0.764$, $0.329$, $0.141$, and $0.061\,\mathrm{pc}$ for L0, L1, L2, and L3, respectively.

Across successively smaller accessible regions, the fitted power law indices increase from $\alpha=1.378\pm0.049$ (L0) to $1.564\pm0.060$ (L1), $1.615\pm0.047$ (L2), and $1.682\pm0.029$ (L3; panel b). This differs from the final L3 index
because of the earlier collapse stage and the different radial fitting interval. Because the fitting intervals differ, this is not a fixed-radius convergence test. Nevertheless, the profiles agree reasonably at overlapping resolved radii, while the inner fits become steeper as smaller radii are reached. Their inward power-law continuations would therefore imply different densities at a common protostellar radius, but these are extrapolations and the profile may change below the resolved range.

A comparison with the fiducial minihalo of \citet{HiranoYoshida2025} reinforces the diversity of possible outcomes. Over the same physical fitting interval, $0.068$--$0.185\,\mathrm{pc}$, Figure~\ref{fig:halom_resolution_comparison}(c) gives $\alpha=1.708\pm0.035$ for HaloM but only $0.443\pm0.054$ for HaloHY. HaloM also has a much higher central dark matter density. The comparison is not strictly matched in evolutionary stage: HaloM is shown at $z=29.53$ and $\rho_{\rm gas,peak}=10^{12}\,\mathrm{GeV\,cm^{-3}}$, whereas HaloHY is at $z=20.65$ and $\rho_{\rm gas,peak}=1.92\times10^9\,\mathrm{GeV\,cm^{-3}}$. The difference therefore cannot be attributed uniquely to halo properties. Nevertheless, the near-overlap of HaloM's late-stage resolved dark matter profiles suggests that its relatively steep structure was already established by the $10^9\,\mathrm{GeV\,cm^{-3}}$ gas-density stage, which is closer to that of HaloHY. The contrast may be related to halo assembly or to the dark matter
orbital distribution, but a quantitative comparison requires profiles and kinematics evaluated at matched collapse stages.

\begin{figure*}[t]
\centering
\includegraphics[width=0.75\textwidth]
{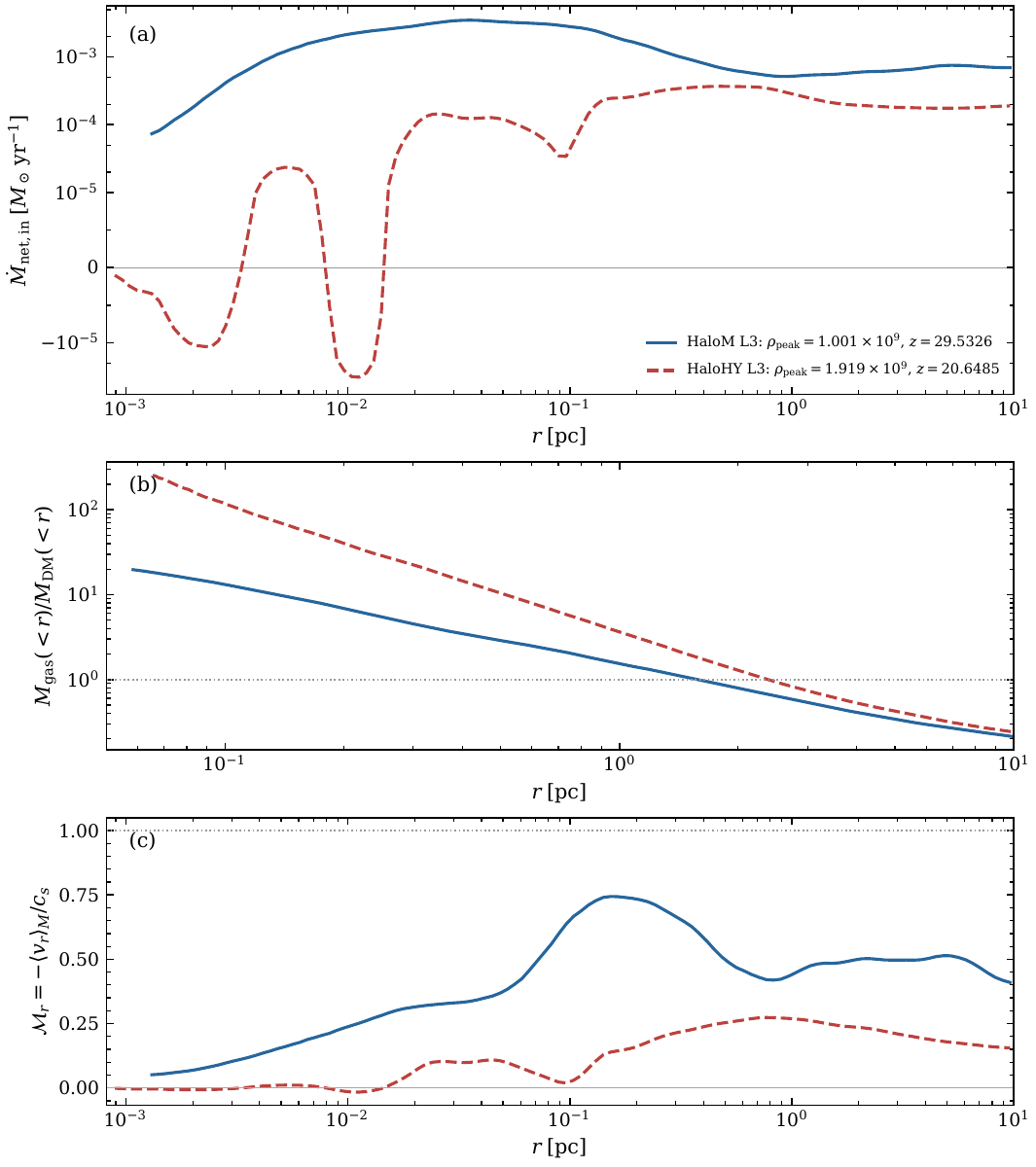}
\caption{
Gas inflow and enclosed-mass comparison between
HaloM L3 (blue solid) and HaloHY
\citep{HiranoYoshida2025} (red dashed) at
approximately matched gas-collapse stages.
The snapshots correspond to peak gas densities
of $1.001\times10^9$ and
$1.919\times10^9\,\mathrm{GeV\,cm^{-3}}$,
at $z=29.5326$ and $20.6485$, respectively.
\textbf{(a)} Net inward gas mass flow rate,
with positive values indicating net inflow.
\textbf{(b)} Ratio of enclosed gas to dark
matter mass.
\textbf{(c)} Mass-weighted radial inflow Mach
number, with positive values indicating mean
inward motion.
These quantities describe the gas before
protostar formation, rather than a measured
protostellar accretion history.
}
\label{fig:halom_hy25_inflow}
\end{figure*}

To compare gas supply at approximately matched collapse stages, Figure~\ref{fig:halom_hy25_inflow} uses peak gas densities of $1.001\times10^9$ (HaloM) and $1.919\times10^9\,\mathrm{GeV\,cm^{-3}}$ (HaloHY). HaloM has a substantially larger net inward gas mass flux and higher inward radial Mach number over most plotted radii. HaloHY nevertheless has a larger enclosed gas-to-dark-matter mass ratio, so relative baryonic dominance alone does not explain the cusp contrast. Stronger instantaneous inflow in HaloM suggests a potentially greater supply of gas for later protostellar growth, but these diagnostics do not establish a sustained accretion history after star formation.

\begin{figure*}[t]
\centering
\includegraphics[width=0.96\textwidth]{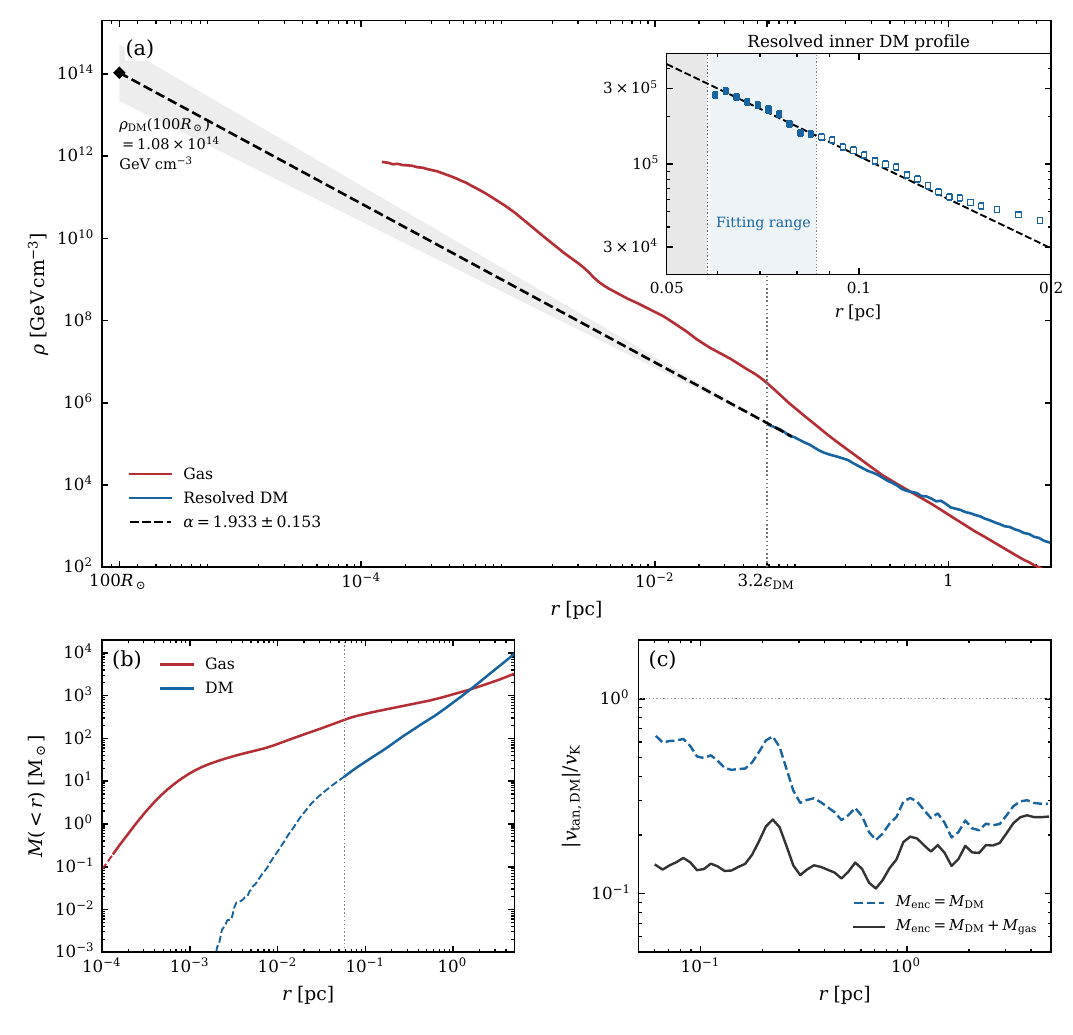}

\caption{Steep inner dark matter cusp and dynamical
structure of HaloM L3 at $z=29.529688$ and peak
gas density $10^{12}\,\mathrm{GeV\,cm^{-3}}$.
\textbf{(a)} Spherically averaged gas (red) and
resolved dark matter (blue) density profiles,
together with the preferred inner power-law fit,
$\alpha=1.933\pm0.153$.
The inset provides a magnified view of the
resolved inner dark matter profile.
The light-blue shaded region indicates the
fitting interval, $r\simeq0.058$--$0.086\,\mathrm{pc}$,
with filled blue symbols denoting the fitted bins
and open symbols showing the remaining bins.
The vertical dotted line in the main panel marks
the adopted softening-based resolution boundary,
$3.2\epsilon_\chi\simeq0.0579\,\mathrm{pc}$.
The black dashed line shows the fitted power law
extrapolated inward to $100\,R_\odot$, where the
black diamond denotes
$\rho_\chi=1.08\times10^{14}\,
\mathrm{GeV\,cm^{-3}}$.
The grey shaded envelope represents the formal
fit uncertainty propagated under the assumption
of an unchanged inner power law.
\textbf{(b)} Enclosed gas and dark matter masses;
the innermost dashed dark matter segment falls
below the adopted dark matter fitting limit.
\textbf{(c)} Dark matter tangential speed divided
by the characteristic Keplerian velocity,
calculated using either the enclosed dark matter
mass alone (blue dashed) or the combined gas and
dark matter mass (dark solid).
Both ratios characterize tangential motion
relative to the local circular-speed scale.}
\label{fig:halom_final_cusp}
\end{figure*}

We examine the final L3 profile more closely in Figure~\ref{fig:halom_final_cusp}. At $z=29.53$, when the peak gas density has reached approximately $10^{12}\,\mathrm{GeV\,cm^{-3}}$, the physical dark matter softening is $\epsilon_\chi\simeq0.01809\,\mathrm{pc}$. The corresponding $3.2\epsilon_\chi$ limit is $0.0579\,\mathrm{pc}$.

The adopted steepest-admissible inner fit spans
$0.058$--$0.086\,\mathrm{pc}$ and contains 10 radial bins, giving
\begin{equation}
{\alpha_{\rm inner}=1.933\pm0.153},\qquad
\rho_\chi(r)\propto r^{-\alpha_{\rm inner}}.
\label{eq:halom_inner_slope}
\end{equation}
To examine radial steepening beyond this selected interval, we also fit three adjacent, non-overlapping 10-bin windows across $0.058$--$0.185\,\mathrm{pc}$, containing $73{,}322$ dark matter particles in total. The middle and outer windows give $\alpha=1.770\pm0.042$ over $0.086$--$0.126\,\mathrm{pc}$ and $1.198\pm0.065$ over $0.126$--$0.185\,\mathrm{pc}$. Thus, the profile steepens inward, although the two innermost indices agree within their formal errors. The inner fit is steeper than the $r^{-1}$ NFW asymptote \citep{Navarro1997}; its uncertainty is the formal fitting error, not a convergence estimate.

The mass profiles in panel (b) show why the gas is dynamically important. Baryons dominate the enclosed mass over much of the inner region, deepening the potential experienced by the dark matter as collapse continues. The dashed innermost portion of the dark matter enclosed-mass profile lies below the adopted fit boundary and is not treated as a converged measurement. Panel (c) compares the characteristic dark matter tangential speed with the local circular-speed scale, $v_{\rm K}(r)=[GM(<r)/r]^{1/2}$, for dark matter mass alone and for the total enclosed mass. Both ratios remain mostly below unity. Unlike the fiducial halo of \citet{HiranoYoshida2025} discussed later, HaloM therefore shows no near-Keplerian core in this shell-averaged tangential-speed diagnostic. This statement does not exclude individual particles on near-circular orbits.

The measured dark matter profile does not reach protostellar radii. To estimate the possible annihilation reservoir, we continue the fitted power law inward to
$100\,R_\odot=2.25\times10^{-6}\,\mathrm{pc}$, obtaining
\begin{equation}
\rho_\chi(100R_\odot)=
1.08\times10^{14}\,\mathrm{GeV\,cm^{-3}}.
\label{eq:halom_extrap_density}
\end{equation}
This radius is approximately $2.6\times10^4$ times smaller than the innermost fitted radius. The inferred density is consequently an extrapolation, conditional on the inner power law persisting to smaller scales. The grey envelope in panel (a) propagates the formal fit uncertainty under that assumption; it does not capture a possible physical break in the profile.

Using this extrapolated DM density, we can estimate the local heating rate due to DM annihilation on scales down to a typical Pop III.1 protostellar radius of $100R_\odot$. For a benchmark dark matter particle mass $m_\chi=500\,\mathrm{GeV}$ and cross section $\langle\sigma v\rangle=3\times10^{-26}\,\mathrm{cm^3\,s^{-1}}$, Equation~(\ref{eq:heating}) then gives
\begin{equation}
\Gamma_\chi(100R_\odot)\simeq
1.12\times10^{-3}f_{\rm heat}\,
\mathrm{erg\,cm^{-3}\,s^{-1}},
\label{eq:halom_illustrative_heating}
\end{equation} 
where $f_{\rm heat}$ is the fraction of annihilation energy assumed to heat the gas locally. For the fiducial $f_{\rm heat}=0.1$,
the estimate is $1.12\times10^{-4}\,\mathrm{erg\,cm^{-3}\,s^{-1}}$. We note that neither transport of annihilation products nor the gas response to this additional heating is included in the calculation.

\begin{figure*}[t!]
\centering
\includegraphics[width=0.90\textwidth]{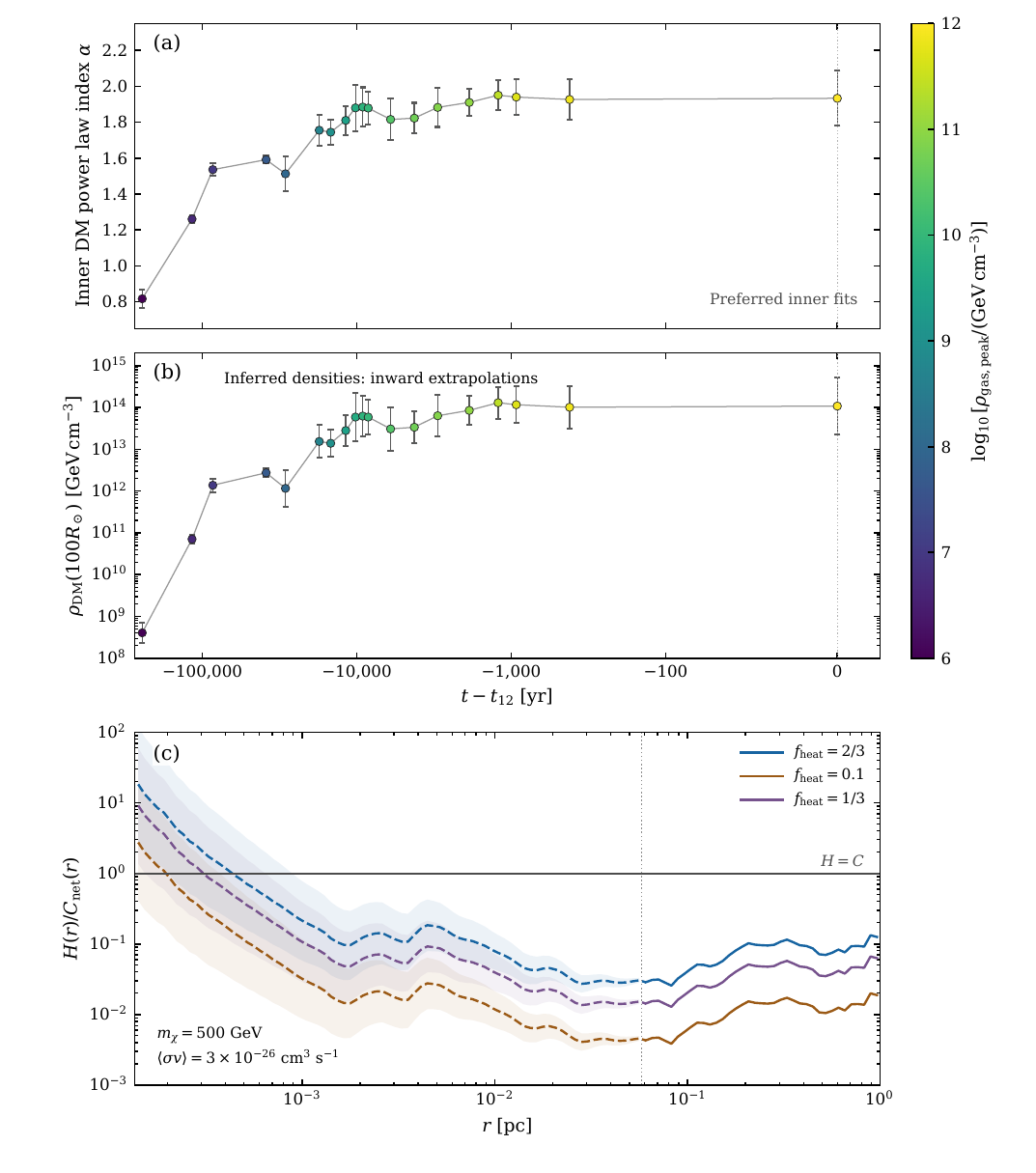}
\caption{Evolution of the inner dark matter cusp in the HaloM L3 simulation and its implications for
annihilation heating.
\textbf{(a)} Evolution of the preferred fitted inner dark matter power law index, $\alpha$, versus cosmological time relative to the final analyzed output, $t_{12}$, for which $\rho_{\rm gas,peak}=10^{12}\,\mathrm{GeV\,cm^{-3}}$.
\textbf{(b)} Dark matter density inferred at $100\,R_\odot$ by extrapolating the fitted inner power law at each L3 output. Marker colors indicate peak gas density as shown by the shared color bar; error bars denote formal fitting uncertainties
(and their propagation into panel (b)).
\textbf{(c)} Ratio of estimated local dark matter annihilation heating, $H(r)\equiv\Gamma_\chi(r)$, to the adopted net gas cooling rate, $C_{\rm net}(r)$, in the final L3 output. Curves correspond to constant deposition fractions $f_{\rm heat}=2/3$, $1/3$, and $0.1$, for $m_\chi=500\,\mathrm{GeV}$ and
$\langle\sigma v\rangle=3\times10^{-26}\,\mathrm{cm^3\,s^{-1}}$. Solid segments use the resolved dark matter profile; dashed segments use its inward power-law extrapolation. The vertical dotted line marks the adopted dark matter resolution boundary, and the horizontal line marks $H=C_{\rm net}$. Shaded envelopes illustrate uncertainties associated with the extrapolated inner cusp. 
}
\label{fig:halom_cusp_time}
\end{figure*}

The time sequence in Figure~\ref{fig:halom_cusp_time} shows the gradual emergence of the steep inner cusp.
Measured relative to the last L3 output, at $\rho_{\rm gas,peak}=10^{12}\,
\mathrm{GeV\,cm^{-3}}$, the preferred inner power law index begins near
$\alpha\simeq0.8$, rises to $1.5$--$1.6$, and eventually approaches
$1.9$ (panel a). Most of this increase occurs before the final high-density outputs. The index subsequently varies little compared with its formal uncertainties, although the gas continues to collapse over the approximately $10^5\,\mathrm{yr}$ shown.

The same development appears in panel (b), where the density inferred at $100\,R_\odot$ rises by several orders of magnitude and reaches approximately $10^{14}\,\mathrm{GeV\,cm^{-3}}$. These values are inward extrapolations of the preferred fitted profiles at each output. Since the fitting interval can vary with time, the series tracks the evolving inner cusp rather than the
index at one fixed physical radius.

Panel (c) asks whether the final cusp, if continued inward, could supply heating comparable to the gas's net microphysical cooling. For $f_{\rm heat}=0.1$, $1/3$, and $2/3$, the estimated local annihilation heating remains below $C_{\rm net}$ over the resolved dark matter radii. In the extrapolated region the ratio rises and can exceed unity at radii of a few $10^{-4}\,\mathrm{pc}$,
with the crossing moving outward as the assumed deposition fraction increases. This identifies a possible WIMP annihilation heating dominated region.

\section{Discussion and Conclusions}
\label{sec:discussion}

The inner dark matter density profile in HaloM steepens during primordial gas collapse and toward smaller resolved radii, reaching $\alpha=1.933\pm0.153$ in the final L3 snapshot. Inward extrapolation of the innermost fit gives $\rho_\chi(100R_\odot)=1.08\times10^{14}\,
\mathrm{GeV\,cm^{-3}}$.

The heating-to-cooling comparison in Figure~\ref{fig:halom_cusp_time}(c) suggests that dark matter annihilation could become thermodynamically important if the fitted cusp continues inward. Such heating could increase thermal pressure, slow central gas collapse, and modify the dark matter response to the evolving potential.

The different inner structures of HaloM and the fiducial minihalo, HaloHY, of \citet{HiranoYoshida2025} may be linked to their orbital support and assembly histories. The nearly unchanged resolved dark matter profiles between the $10^9$ and $10^{12}\,\mathrm{GeV\,cm^{-3}}$ stages may reflect the decreasing gas free-fall time relative to dark matter orbital timescales, or continued contraction at smaller, unresolved radii. In HaloHY, a near- or super-Keplerian dark matter core initially inhibits inward motion, before continued baryonic infall drives renewed contraction. In contrast, the characteristic tangential speed of HaloM's inner dark matter remains below the local Keplerian speed, whether this is calculated from dark matter alone or from the combined mass of gas and dark matter (Fig.~\ref{fig:halom_final_cusp}c). Thus, unlike in HaloHY, our shell-averaged profiles do not show a near-Keplerian dark matter core capable of providing the same angular-momentum barrier. This may allow lower-angular-momentum orbits to reach the baryon-dominated region and contribute to the steep cusp. Low global halo spin may provide a useful selection criterion in an $N$-body catalogue, although it does not directly measure inner rotational support.

At approximately matched gas-collapse stages, Figure~\ref{fig:halom_hy25_inflow} also shows stronger net inward gas transport and larger sound-speed-normalized radial inflow in HaloM than in HaloHY, despite the latter's larger enclosed gas-to-dark-matter mass ratio. The coexistence of a steep dark matter cusp and strong gas inflow is particularly favorable for massive Pop III.1 star formation if the inflow is sustained. However, at accretion rates of $\sim10^{-3}\:M_\odot\:{\rm yr}^{-1}$ it would still take $\sim10^7\:$yr for the star to grow to $\sim10^4\:M_\odot$.

The present simulation ends before the formation and early evolution of a protostar can be followed. This is important when comparing our fiducial extrapolation, $\rho_\chi(100R_\odot)=1.08\times10^{14}\, \mathrm{GeV\,cm^{-3}}$, with the ambient dark matter densities of order $5\times10^{14}\,\mathrm{GeV\,cm^{-3}}$ associated with supermassive growth in the stellar-evolution models of \citet{Nandal2025} and \citet{Topalakis2025}. The two quantities are not directly equivalent: the stellar models specify an ambient dark matter density for a WIMP capture model, whereas our value is a radius-specific extrapolation from an earlier gas-collapse snapshot. Our estimate lies below that illustrative benchmark, but it does not determine the density that would surround a protostar at the
later stage. Two distinct trends motivate a denser inner cusp at later stages: the L0--L3 sequence reveals steeper fits as progressively smaller
radii become accessible, while the L3 time sequence shows the cusp steepening as the gas collapses. The corresponding inward extrapolations are sensitive to both slope and normalization, and favor higher
protostellar-scale dark matter densities if the steepening persists. For illustration, if the density at $0.058\,\mathrm{pc}$ were held fixed, an additional steepening of $\Delta\alpha\simeq0.15$ sustained down to $100\,R_\odot$ would raise the extrapolated density by a factor of about $4.6$. This is a sensitivity estimate, not a prediction that such steepening must occur.

Higher-resolution calculations and tests of gravitational softening and particle mass \citep{Power2003,Hopkins2018,Ludlow2020}, together
with the inclusion of annihilation heating, will help determine the inner cusp structure and its impact on subsequent gas accretion and protostellar evolution.

We conclude that primordial gas collapse in a Pop III.1 minihalo can be accompanied by a steep, time-dependent dark matter cusp and by a strong pre-protostellar gas inflow. Their coexistence motivates further tests of the unresolved annihilation reservoir and whether later WIMP capture and gas accretion can sustain the growth of
supermassive Pop III.1 stars as progenitors of SMBHs.

\begin{acknowledgments}
The computations presented in this work were performed on the Dardel supercomputer at the PDC Center for High Performance Computing, KTH Royal Institute of Technology, Sweden. Computational resources were provided by the National Academic Infrastructure for Supercomputing in Sweden (NAISS), funded by the Swedish Research Council.
S.B. acknowledges support from a Chalmers Astrophysics and Space Sciences Summer (CASSUM) fellowship. J.C.T. acknowledges support from the Chalmers Initiative on Cosmic Origins (CICO) and the Virginia Institute for Theoretical Astrophysics (VITA), supported by the College and Graduate School of Arts and Sciences at the University of Virginia.
S.H. acknowledges support from JSPS KAKENHI grants JP21K13960 and JP26K00743.
\end{acknowledgments}

\bibliographystyle{journalbib}
\bibliography{reference}

\end{document}